\documentclass[12pt]{article}
\usepackage{amsmath,amsthm}
\theoremstyle{definition}
\newtheorem{example}{Example}
\usepackage{graphicx}
\usepackage[round]{natbib} 
\usepackage{url} 
	\usepackage{easyReview}
\usepackage[linesnumbered,ruled]{algorithm2e}
\newcommand{\blind}{0}

\begin{document}

\def\spacingset#1{\renewcommand{\baselinestretch}%
{#1}\small\normalsize} \spacingset{1}


\if0\blind
{
  \title{\bf Factor selection in screening experiments by aggregation over random models}
  \author{Rakhi Singh\\
    
    University of North Carolina at Greensboro, Greensboro, NC, USA\\
    and \\
    John Stufken\thanks{
    The authors gratefully acknowledge support through NSF grant DMS-1935729}\hspace{.2cm} \\
    University of North Carolina at Greensboro, Greensboro, NC, USA}
  \maketitle
} \fi

\if1\blind
{
  \bigskip
  \bigskip
  \bigskip
  \begin{center}
    {\LARGE\bf Factor selection in screening experiments by aggregation over random models}
\end{center}
  \medskip
} \fi

\bigskip
\begin{abstract}
Screening experiments are useful for screening out a small number of truly important factors from a large number of potentially important factors. The Gauss-Dantzig Selector (GDS) is often the preferred analysis method for screening experiments. Just considering main-effects models can result in erroneous conclusions, but including interaction terms, even if restricted to two-factor interactions, increases the number of model terms dramatically and challenges the GDS analysis. We propose a new analysis method, called Gauss-Dantzig Selector Aggregation over Random Models (GDS-ARM), which performs a GDS analysis on multiple models that include only some randomly selected interactions. Results from these different analyses are then aggregated to identify the important factors. We discuss the proposed method, suggest choices for the tuning parameters, and study its performance on real and simulated data. 
\end{abstract}

\noindent%
\vfill

\newpage
\spacingset{2} 
\section{Introduction}
\label{sec:intro}
With a large number of potentially important factors, screening experiments are an economical choice for screening out the truly important factors. Such an experiment would then often be followed by a second experiment to study the effects of the selected factors in more detail. 

Factors can be important through main-effects or interactions. Hence, with a large number of factors, there are many model terms to consider. Coupled with a small number of runs, it is common to assume that only a small number of effects are active (effect sparsity). Throughout, we make this assumption. The effects deemed active by the model selection method are used to select factors for a follow-up experiment. It is not uncommon to use two levels for each factor, although a qualitative factor may naturally have three or more levels. We restrict consideration to designs with $n$ runs and $m$ two-level factors, although the proposed methodology can be extended to factors with more than two levels. The number of runs $n$ is often small, and less than the number of model parameters to be estimated. A design used for such an experiment is said to be supersaturated.

With a large number of factors, the process to be investigated is typically not well understood, so that one cannot assume that a main-effects model will be adequate. Based on the effect hierarchy principle, we restrict attention to main-effects and two-factor interactions. With $m$ 2-level factors, there are $m + {m \choose 2}$ such terms. For $m=15$, say, there are already 120 terms. Screening focuses on identifying important factors. A factor is important if it is part of an effect that is found to be active. Since we only consider main-effects and two-factor interactions, a factor is important if its main-effect is active or if it is involved in an active two-factor interaction. With $n=20$ observations, say, and 120 model terms, this is a very complex problem. 

We will see that there is a need for a new method of analysis for problems with this level of complexity. We will introduce this method in Section~\ref{newmethod}. In the remainder of this section, we provide a brief introduction to designs, existing methods of analysis, and a first peek at the proposed method. 
 
\subsection{Designs}\label{sec-1.1}
For a main-effects model, a 2-level design is supersaturated if $n<1+m$. The literature on optimality criteria and design construction for such supersaturated designs is vast  \citep[see,][for example]{BC62,jones2008,JM14,ryanGOSSD, mariaweese2020}. For a consolidated review, we refer to \cite{georgiou}, and to \cite{mariaweese2020} for more recent developments. Among the studied optimality criteria are $E(s^2)$, $UE(s^2)$, Bayesian D-optimality and Var($s+$). However, these criteria are primarily used for main-effects models in situations where $n<1+m$. Constructions of supersaturated designs allowing the identification of active main-effects without the assumption that two-factor interactions are negligible have been studied in \cite{shitang}. 

A 2-level design is supersaturated for a model with main-effects and two-factor interactions if $n< 1+m + {m \choose 2}$. While $n$ may still be less than $1+m$, designs with larger runs will be much preferred if resources are available. For example, in \cite{Draguljic2014}, for screening $m=10$, 15, and 20 factors, Bayesian D-optimal designs with $n=32$, 58, and 94 runs, respectively, are used. Based on the simulations in \cite{Draguljic2014}, these designs perform well even with multiple two-factor interactions. But it may not always be feasible to use designs of this size. More sophisticated methods such as the mixed-integer optimization approach studied in \cite{vazquezetal2020} could also be useful to impose user-defined restrictions (for instance, imposing effect heredity, treating multi-level categorical factors as a group, setting bounds on model size).

\subsection{Existing analysis methods}
Multiple analysis methods have been studied for screening experiments. This includes popular methods with much broader applications, such as forward selection \citep{westfalletal}, LASSO \citep{tibs96}, smoothly clipped absolute deviation, SCAD \citep{fanli2001}, the Gauss Dantzig selector, GDS \citep{CandesTao}, and simulated annealing model search, SAMS \citep{wolters}. Comparisons between most of these have been made for screening experiments.  

For example, \cite{phoa2009} adapted the GDS for supersaturated designs. Subsequently, \cite{MW10} and \cite{mariaweese2015,mariaweese2017} showed that, for a main-effects model, among forward selection, GDS, and model averaging, GDS performs better for recovering active effects along with making a limited number of errors. For a model with main-effects and two-factor interactions, GDS has also been identified as the best screening method among SCAD, LASSO, GDS, SAMS, Bayesian model selection, and Bayesian model averaging in \cite{Draguljic2014}. Group screening methods discussed in \cite{Draguljic2014} and \cite{ryanGOSSD} can also be considered, but may require a large number of runs or prior knowledge about potentially active effects.

In the context of observational studies, \cite{bienetal} propose an extension of LASSO, called hierarchical LASSO, for studying models with two-factor interactions. Models are assumed to satisfy the effect heredity principle \citep{hamadawu92}, meaning that a two-factor interaction can only be active if at least one of the corresponding main-effects is active. The authors also study the effect of applying LASSO on models with all main-effects and two-factor interactions (called \textit{all-pairs LASSO}) and conclude that hierarchical LASSO works better than all-pairs LASSO. Two additional analysis methods have been studied in \cite{phoachemo}: the frequentist approach of \cite{hamadawu92} and the Bayesian approach of \cite{boxmeyer}. These methods can be computationally expensive for larger values of $m$. The first method also uses the forward selection regression procedure. As mentioned before, \cite{MW10} concluded, albeit for main-effects models, that GDS performs better than forward selection. 


For the reasons listed, we develop our method based on GDS. While other methods could have been used in principle, a separate study would be needed for comparison and tuning method-specific parameters.

\subsection{A first peek at the new method} 
GDS can be applied using only the $m$ main-effects or using the $m$ main-effects and all ${m \choose 2}$ two-factor interactions. We will refer to these as GDS(m) and GDS(m+2fi), respectively. The latter is akin to \textit{all-pairs LASSO}. We will see that both methods are unsatisfactory. Instead, we propose to apply GDS many times, each time with all main-effects and a randomly selected set of two-factor interactions. We then use a method of aggregation over the models suggested by the GDS applications to select the potentially active effects. As a final step, we apply a stepwise regression starting from the model with the selected effects, and giving main-effects that were not selected an opportunity to enter the model. We call this method \textit{GDS-ARM} (Gauss Dantzig Selector--Aggregation over Random Models). GDS-ARM gives consideration to two-factor interactions, while reducing the complexity faced by GDS(m+2fi) when all two-factor interactions are considered simultaneously. Akin to random forests, it does so by each time using only some of the two-factor interactions. An assessment of the performance of GDS-ARM, along with choices for tuning parameters, will be provided in the next sections.



\section{Preliminaries}\label{sec:background}
A common model for screening experiments is
\begin{equation}
y= X\beta+\epsilon, 
\label{eq-model}
\end{equation}
where $y$ and $\epsilon$ are $n\times 1$ vectors of responses and errors, respectively, $X$ is the $n\times p$ model matrix, and $\beta$ the $p\times 1$ vector of parameters. For 2-level factors, the entries of $X$ would all be $\pm 1$. But we center $y$ and center and normalize the columns of $X$ (to the length $\sqrt{(n-1)}$), calling the resulting vector and matrix again $y$ and $X$, respectively. As a result, we do not need an intercept parameter in \eqref{eq-model}, and take $p = m$ or $p = m + {m \choose 2}$ for a main-effects model and a model that also includes all two-factor interactions, respectively.

\subsection{The Gauss-Dantzig selector \label{subsec-GDS}}
The GDS was first proposed by \cite{CandesTao} in the multiple regression context. It starts with the Dantzig selector, which obtains the estimator $\hat{\beta}$ as a solution to 
\begin{equation}
\min_{\beta \in {\mathcal R}^p} ||\beta||_1 \hspace{0.5cm} \mbox{ subject to  }\hspace{0.4cm} ||X^T(y-X\beta)||_{\infty} \le \delta, 
\label{eq-GDS}
\end{equation}
where $||\beta||_1= |\beta_1|+\cdots+|\beta_{p}|$ is the $l_1$ norm, and $||a||_{\infty}= \max(|a_1|,\dots, |a_{p}|)$ is the $l_{\infty}$ norm. As a second step, only the effects with estimates exceeding $\gamma$ in magnitude are retained, and are re-estimated using ordinary least squares. The second step is helpful in reducing the bias of the estimates. The whole process is repeated for multiple values of $\delta \in (0, ||X^Ty||_{\infty})$, leading to multiple models. Model selection criteria like AIC, BIC, adjusted $R^2$, etc. can then be used to select one of these models (see \cite{phoa2009} and \cite{MW10}). There is no consensus on how to select a value for $\gamma$. \cite{phoa2009} suggest using a data-driven value such as $0.1 \times ||\hat\beta||_{\infty}$, while \cite{MW10} use $\gamma=1.5$ in their simulations. Recently, \cite{mariaweese2020} evaluate three different choices for $\gamma$ in their simulatons, and also suggest the data-driven value of $0.1 \times ||\hat\beta||_{\infty}$. A value of $\gamma$ that is too small results in models that are too large, but if $\gamma$ is too large we may miss active effects. While we concur that a data-driven selection method is preferred, we have seen too many examples where $0.1 \times ||\hat\beta||_{\infty}$ is not a good choice. In Subsection \ref{subsec-gamma}, we provide an alternative method which works reasonably well across simulated and real datasets.

\subsection{A closer look at GDS(m) and GDS(m+2fi)}
GDS(m) is prone to two types of errors. First, it may miss an important factor if that factor is only active through an interaction.  Second, it may incorrectly identify a factor as important if the main-effect for that factor is highly correlated with an active two-factor interaction. To illustrate this, we discuss the following example based on a real experiment.


\begin{example}\label{exWuHamada} 
A cast fatigue experiment with 12 runs and 7 factors was originally studied by \cite{Hunteretal}, and was later revisited by \cite{hamadawu92} and \cite{phoa2009}, among others. For ease of reference, the data are provided in the Supplementary Material. Table \ref{tab-resultsWuHamada} shows the active effects identified by GDS(m) and GDS(m+2fi) for different choices of tuning parameters. For tuning $\delta$, the results are identical for AIC or BIC, but the choice of $\gamma$ makes a difference. It is widely accepted for these data that F and FG are active effects, with AE possibly being active as well (see \cite{hamadawu92}). Yet, GDS(m) only identifies F correctly, while incorrectly declaring D to be important. The estimate for the main-effect of $G$ is the second smallest in magnitude when using the Dantzig selector with a main-effects model. Thus, if G would be declared an important factor with GDS(m), then four other factors would incorrectly be identified as important. Note that GDS(m+2fi) works well for this example, but not for the suggested choice of $0.1 \times ||\hat\beta||_{\infty}$ for $\gamma$. The $R^2$ values for the various models are 59\% for D and F, 89\% for F and FG, and 95\% for F, FG and AE.

\begin{table}[hbtp]
\caption{GDS results on the cast fatigue experiment in Example \ref{exWuHamada}}
\begin{center}
\begin{tabular}{ccc}
\hline\noalign{\smallskip}
Method &$\gamma$ & Active Effects \\
\noalign{\smallskip}\hline\noalign{\smallskip}
GDS(m) &  $0.1 \times ||\hat\beta||_{\infty}$ & D, F \\\noalign{\smallskip}\hline
GDS(m) & $0.4 \times ||\hat\beta||_{\infty}$ & D, F \\\noalign{\smallskip}\hline
GDS(m) & Clustering (S. \ref{subsec-gamma}) & D, F \\\noalign{\smallskip}\hline
GDS(m+2fi) &  $0.1 \times ||\hat\beta||_{\infty}$ & D, F, AD, \\
 &  &  AE, EF, FG \\\noalign{\smallskip}\hline
GDS(m+2fi) & $0.4 \times ||\hat\beta||_{\infty}$ & F, AE, FG \\\noalign{\smallskip}\hline
GDS(m+2fi) &  Clustering (S. \ref{subsec-gamma}) & F, AE, FG \\
\noalign{\smallskip}\hline
\end{tabular}
\end{center}
\label{tab-resultsWuHamada}
\end{table}
\end{example}
In simulations, screening performance is often assessed on two measures: (i) average power, (ii) average (type 1) error, defined as:  
\begin{itemize}
\item Power: the proportion of important factors correctly identified over multiple iterations;
\item Error: the proportion of unimportant factors declared to be important over multiple iterations.
\end{itemize}

Increasing power also tends to increase error. A method should have high power because missing an important factor at the screening stage is highly undesirable. But if the error is too large, too many unimportant factors will be kept for a follow-up experiment. With a slightly higher emphasis placed on higher power than on smaller error, we consider a method to be good if it has higher power and reasonably small error.  As we will see, GDS(m) performs well in terms of power and error if there are no active interactions. We will also see that, irrespective of the presence of active interactions, the performance of GDS(m+2fi) can be poor, especially if $n$ is relatively small. The poor performance of GDS(m+2fi) can primarily be attributed to the fact that GDS needs to consider quite a few columns with some of them being highly correlated. For example, for $n=16, m=24$, GDS(m+2fi) has the sheer impossible task of considering $24+ {24 \choose 2} = 300$ highly correlated potential effects in just 16 runs.

\subsection{An alternative for tuning $\gamma$\label{subsec-gamma}}
The data-driven choice of $\gamma= 0.1 \times ||\hat\beta||_{\infty}$ may work poorly, as in Example~\ref{exWuHamada} for GDS(m+2fi). While $0.4 \times ||\hat\beta||_{\infty}$ worked well in that example, it tends to be too large in many cases. We use an alternative procedure to select active effects throughout the paper:
\begin{enumerate}
\item For selected values of $\delta$, obtain the Dantzig selector estimate, say $\hat\beta (\delta)$, of $\beta$.
\item For each $\hat\beta (\delta)$, apply k-means clustering \citep{lloyd82} with two clusters on the absolute values of the coefficients in $\hat\beta (\delta)$, and, using ordinary least squares, refit a model that only contains effects corresponding to the cluster with the larger mean.
\item Select the value of $\delta$ that corresponds to the model from the previous step that has the smallest BIC value. The effects in that model are declared to be the active effects.

%
\end{enumerate}

We use 10 values for $\delta$ in this process, which form, together with 0 and $||X^Ty||_{\infty}$, 12 equidistant points in $[0,||X^Ty||_{\infty}]$.

This clustering method does not explicitly select a value for $\gamma$, but has the same effect of setting coefficients with smaller estimates equal to 0. Based on our experience, it works well on many simulated and real datasets, including in Example~\ref{exWuHamada}.


\section{The GDS-ARM method} \label{newmethod}
In this section, we describe the GDS-ARM method in more detail and recommend choices for its tuning parameters. Its performance will be studied in Section~\ref{sec-results}.

\subsection{The algorithm}
Schematically, GDS-ARM consists of four steps:
\begin{itemize}
\item[(a)] apply GDS on multiple models with all main-effects and randomly selected interaction columns.
\item[(b)] using a model selection criterion, identify the top models from those obtained in part (a).
\item[(c)] select effects by aggregating over the selected top models in part (b).
\item[(d)] apply stepwise regression starting with a model that contains the effects selected in part (c), and give main-effects that were left out an opportunity to enter the final model.
\end{itemize}    
Effects that appear in the model selected in step (d) are declared to be active, and any factor that is part of at least one such effect is declared to be important. 

For step (a) (lines 1--5 in Algorithm~\ref{algo1}), GDS is performed $n{rep}$ times, each time with $nint$ randomly selected two-factor interactions. Thus, the model matrix $X$ will each time be an $n\times (m+nint)$ matrix, say $X_{m,nint}$, with $m$ main-effects columns and $nint$ interaction columns. Each time, we use the procedure described in Section~\ref{subsec-gamma}. One could try to select the set of interaction columns more judiciously, for example by using balanced incomplete block designs, but randomly selected columns appear to work well and impose no restrictions on the values for $nrep$ and $nint$. 

In step (b) (line 6 in Algorithm~\ref{algo1}), we select $n{top}$ models using the BIC criterion following \cite{MW10} and \cite{mariaweese2020}. Other choices are possible, such as modified AIC \citep{phoa2009} or corrected AIC \citep{Draguljic2014}, however, these choices often result in very small models \citep[see, for example][]{MW10}.

Step (c) (line 7 in Algorithm~\ref{algo1}) discards effects appearing in less than $pkeep \times ntop$ of the top models, where $pkeep$ is the proportion of top models in which an effect needs to appear in order to be declared active. This eliminates effects that appear, perhaps by chance, in only very few top models. 


Finally, with step (d) (line 8 of Algorithm~\ref{algo1}), incorrectly selected effects can exit the model, while main-effects have another chance to enter the model. This preferential treatment for main-effects aligns with the effect hierarchy principle.  In principle, one could use GDS instead of the stepwise regression. We use stepwise because, unlike GDS, it gives us the flexibility to start with a model which we already deem appropriate and with a chance to add additional effects that could potentially be important.

\begin{algorithm}[hbtp]
	\caption{GDS-ARM}\label{algo1}
	\SetKwInOut{Input}{inputs}
	\SetKwInOut{Output}{output}
	\Input{the number of repetitions $n{rep}$, the number of selected two-factor interactions $nint$, the number of top models $n{top}$, the minimum proportion of top models in which any effect should appear $pkeep$, a design $d$, and the response vector $y$}
   	\For{$ j=1 \rightarrow n{rep} $}{
   	   Let $X_{m,nint}$ be a $n\times (m+nint)$ matrix with all rows of $X$, all $m$ main-effects columns, and $nint$ randomly selected interaction columns\;
   	   Apply the GDS procedure in Section \ref{subsec-gamma} using $y$ and $X_{m,nint}$ \;
       Save the selected active effects in the list $B_j$\;
}
	Using BIC, find the best $n{top}$ models (call them, say, $\{B_{(1)},\dots, B_{(n{top})}\}$) from the $n{rep}$ models corresponding to $B_1$, \dots, $B_{nint}$\;
	Select the effects from $\{B_{(1)},\dots, B_{(n{top})}\}$ that appear in at least $\lceil pkeep \times n{top}\rceil$ of these lists\;
	Starting with the model that contains all selected effects, perform stepwise regression on selected effects and all $m$ main-effects\;
	Selected effects from the stepwise regression are declared to be active effects \;
	Declare a factor to be important if it appears in at least one active effect\;
	\Output{important factors from step 10} 
\end{algorithm}

The basic idea and motivation for GDS-ARM comes from random forests. With random forests, instead of using just one tree, multiple trees are used, just as we use multiple GDS applications. And instead of using all variables to build a tree, only randomly selected variables are used each time. As a result, GDS-ARM facilitates inclusion of two-factor interactions while avoiding the complexity that GDS(m+2fi) would face by using all these variables simultaneously.

\subsection{Parameter tuning}
Aside from $\delta$ and $\gamma$ (see Section~\ref{subsec-gamma}), there are four tuning parameters for Algorithm~\ref{algo1}: $n{rep}$, $nint$, $n{top}$, $pkeep$.  

For each of the $nrep$ GDS applications, we use $nint$ two-factor interactions. Since there are $m \choose 2$ such interactions, there are ${{m \choose 2} \choose nint}$ different models that we could fit. This number will almost always be too large. Moreover, while $nrep$ should not be taken too small, we find that there is a point of diminishing returns. For $nint$, note that GDS(m) and GDS(m+2fi) correspond to $nint=0$ and $nint={m \choose 2}$, respectively. We find that a much smaller value for $nint$ than ${m \choose 2}$ is preferable, although the best choice can depend on (1) the number of runs $n$ and (2) the number of active two-factor interactions. With larger values for these numbers (albeit that the number in (2) can only be guessed, such as by using some guidance from \cite{li2006regularities}), one could consider taking $nint$ larger than our recommendation.

The number of top models, $n{top}$, is bounded by $n{rep}$. Some of the fitted models may not provide a good fit because the randomly selected interactions for that GDS application did not include one or more active interactions. On average, an interaction is selected in $n{rep}\times nint/{m \choose 2}$ GDS applications. To give active interactions a good chance to show up in the top models, we would want to take $ntop$ smaller than this number. Finally, for an effect not to be discarded at this stage, we would want it to appear in at least $100 \times pkeep$\% of the top models. Effects that appear in very few of the top models may have appeared there by chance. On the other hand, we would not want $pkeep$ to be too large, because active interactions could appear in only a modest number of the top models.   


\begin{table}[hbtp]
\caption{Parameter tuning for GDS-ARM}

\begin{center}
\begin{tabular}{ccccc}
\hline\noalign{\smallskip}
Param. & Description & Levels & Values& Selected value \\
\noalign{\smallskip}\hline\noalign{\smallskip}
$nrep$ & Number of GDS applications &3 & $\frac{1}{2}{m \choose 2}, {m \choose 2}, 2{m \choose 2}$ & ${m \choose 2}$\\[0.5em]\hline
$nint$ & Number of 2fi in each GDS application &2 & $0.2{m \choose 2}$, $0.4{m \choose 2}$&$0.2{m \choose 2}$\\[0.5em]\hline
$ntop$ & Number of top models &2&$\max\big(20,\frac{n{rep}\times nint}{2{m \choose 2}}\big)$, &$\max\big(20,\frac{n{rep}\times nint}{2{m \choose 2}}\big)$\\[0.75em]
&&&$\frac{2n{rep}\times nint}{{m \choose 2}}$\\[0.75em]\hline
$pkeep$ &Prop. of $ntop$ models in which &3 &$0.10$, $0.25$, $0.4$ & 0.25\\[0.25em]
 &an effect should appear & &&\\[0.25em]
\noalign{\smallskip}\hline
\end{tabular}
\end{center}
\label{tab-partuning}
\end{table}

%
We performed a simulation study using a complete factorial for the four tuning parameters with the 36 level combinations shown in Table~\ref{tab-partuning}. We studied power and error across simulation scenarios S1-S7 described in Subsection~\ref{sec4.2}, using 100 iterations for each level combination and for one design for each of $(n,m)$ = (12, 16), (18, 22), (20, 16), (20, 24), (24, 20) and (24,16). Detailed results can be found in the Supplementary Material. For each of the seven scenarios, we perform an analysis of variance with main-effects and two-factor interactions, using the difference between power and corresponding error as the response. This resulted in the recommendations provided in the final column of Table~\ref{tab-partuning}. 

Thresholds for the p-values in the stepwise procedures in steps 8 and 9 of Algorithm \ref{algo1} are additional tuning parameters. Throughout our real and simulated studies, we set the threshold for adding a term at 0.01 and for removing a term at 0.05. As with other tuning parameters, sometimes other choices for the tuning parameters may be better.

\section{Screening performance\label{sec-results}}
We apply GDS-ARM to two real datasets and to various simulated datasets. Throughout this section, for every GDS application, we use the method provided in Section \ref{subsec-gamma}.

\subsection{Real case studies}
First, we apply GDS-ARM, with values for the tuning parameters as in the last column of Table~\ref{tab-partuning}, to the experiment from Example~\ref{exWuHamada}. 

\noindent\textbf{Example~\ref{exWuHamada} revisited.} 
For the cast fatigue experiment with 12 runs, 7 factors and 28 effects, we saw that GDS(m) incorrectly identifies factors D and F as important, whereas GDS(m+2fi) identifies factors $A$, $E$, $F$ and $G$ (Table~\ref{tab-resultsWuHamada}). The application of GDS-ARM with $nrep= 21$, $nint =5$, $ntop=20$ and $pkeep=0.25$ (from Table~\ref{tab-partuning}) also identifies $A$, $E$, $F$ and $G$ as important factors. 

The next case study is about an analytical experiment conducted by \cite{dopico07} to characterize the chemical composition of white ``Vinho Verde" grapes simultaneously determining the most important phenolic compounds and organic acids for the grapes. We follow \cite{phoachemo} by considering one phenolic compound, kaempferol-3-Orutinoside + isorhamnetin-3-O glucoside. The data is provided in the Supplementary Material for the sake of completeness. 

\begin{example}\label{ex-CompEtxr} The experiment used a 12-run Placket-Burman design with eight factors ($A$--$H$). Fitting a main-effects model suggests that factors $D$ and $F$ are important (the corresponding model has $R^2=41\%$). Upon reanalyzing the data, \cite{phoachemo} found that the active effects are $C$, $D$, and $AD$ ($R^2=93\%$), so that the important factors are $A$, $C$, and $D$. They also note that the factor $F$ was misidentified in the main-effects analysis perhaps due to its partial aliasing with the interaction $AD$. From Table~\ref{tab-resultsCompExtr}, we see that both GDS(m) and GDS(m+2fi) misidentify important factors, whereas GDS-ARM with parameters as in Table~\ref{tab-partuning} reaches the same conclusion as in \cite{phoachemo}. 

\begin{table}[hbtp]
\caption{GDS results on the chemical composition experiment in Example \ref{ex-CompEtxr}}
\begin{center}
\begin{tabular}{lll}
\hline\noalign{\smallskip}
Method & Parameters & Imp. factors \\
\noalign{\smallskip}\hline\noalign{\smallskip}
GDS(m) & BIC, Clustering & B, D, E, F \\
GDS(m+2fi) & BIC, Clustering & A, B, D, E, G, H \\
GDS-ARM &$nrep= 28, nint =6,$ & A, C, D \\
 &$ntop=20, pkeep=.25$ & \\
\noalign{\smallskip}\hline
\end{tabular}
\end{center}
\label{tab-resultsCompExtr}
\end{table}

\end{example}

\subsection{Simulated datasets} \label{sec4.2}
For given numbers of $n$ runs and $m$ factors, we consider seven scenarios, say S1 through S7. The numbers of active main-effects, $c_1$, and two-factor interactions, $c_2$, are taken as $(c_1,c_2)= (3,0), (4,0), (5,0)$, $(3,1)$, $(4,1)$, $(3,2)$ and $(4,2)$ for S1 through S7, respectively. Unless otherwise stated, two-factor interactions can only be active if at least one of the corresponding main-effects is active (weak heredity). Hence, there are at most $c_1+c_2$ important factors. GDS-ARM is expected to do even better compared to GDS(m) if there are active interactions between factors whose main-effects are not active, even though such situations are not considered here. So, scenarios S1-S3 have 0 active two-factor interactions (which favors GDS(m)), while the other scenarios have 1 or 2 active two-factor interactions. All coefficients for active effects (including the intercept) are generated from $N(5,1)$, with a randomly selected sign. Coefficients corresponding to inactive effects are set to 0. Errors are generated as independent $N(0,1)$. Thus, unless otherwise stated, the mean effect sizes in our simulations are 5. In particular, in Example \ref{exmpln18m22} and the Supplementary Material (for Example \ref{exmpln24m16}), we consider mean effect sizes of 5, 3 and 1.5. Using a design $d$, and corresponding design matrix $X$ (prior to centering and normalizing $X$), the response is then generated using  \eqref{eq-model}. For each scenario and design, a response vector is generated 1000 times. Average screening performance over the 1000 iterations is measured by power and error. 

We present results for $(n,m)$ = $(18,22)$ and $(24,16)$ and show that GDS-ARM is superior to GDS(m) and GDS(m+2fi). In Example~\ref{exn14to20m24}, we consider performances of designs with varying $n=14-20$ and $m =24$. These combinations of $(n,m)$ have been used in previous studies and designs used have different properties. While we studied multiple designs for each case, since performance is similar, results for only one design in each case are presented, with those for other designs being relegated to the Supplementary Material. For GDS(m) and GDS(m+2fi), we use BIC and the clustering-based method for parameter tuning. While use of the adaptive value of $\gamma = 0.1  \times ||\hat\beta||_{\infty}$ would have resulted in slightly larger power, it would also have resulted in considerably larger error. For GDS-ARM we use the parameters suggested in Table~\ref{tab-partuning} and the methodology described in Section~\ref{subsec-gamma}. In all figures, we plot power (solid lines) and error (dashed lines).

\begin{figure}[hbtp]
	\centering
		\includegraphics[scale=0.6]{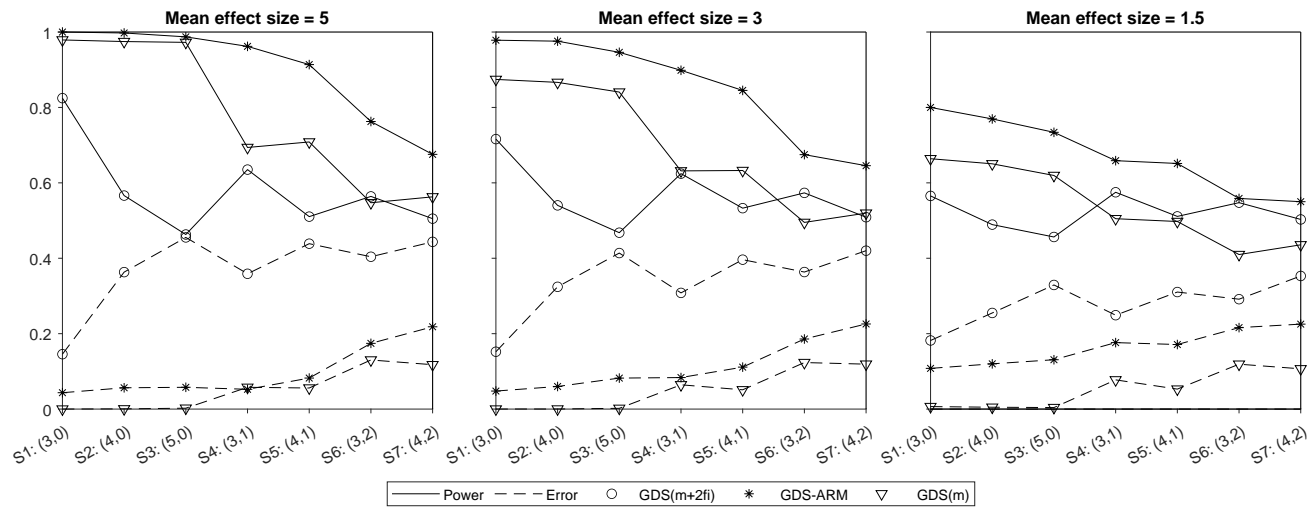}
	\vspace*{-10mm}
	\caption{Average power and error over 1000 iterations for $n=18,m=22$ for an E$(s^2)$-optimal design when coefficients of active effects are generated from a normal distribution with standard deviation 1 and mean effect sizes of 5 (left panel), 3 (middle panel), and 1.5 (right panel).}
	\label{fig-Main-n18m22}
\end{figure}

\begin{example}\label{exmpln18m22}
For $n=18$ and $m=22$, we use the $E(s^2)$-optimal design in \cite{MW10}. The tuning parameters for GDS-ARM are $n{rep}=237$, $nint=47$, $n{top} = 24$, and $pkeep=0.25$. The mean effect sizes of 5, 3, and 1.5 for the coefficients of active effects are considered in the left, middle, and right panels of Figure~\ref{fig-Main-n18m22}. We have ensured that for an active interaction at least one of the factors has an active main effect. The left most panel of Figure~\ref{fig-Main-n18m22} shows that GDS(m+2fi) performs poorly in terms of both power and error. GDS(m) performs well for the scenarios without interactions (S1-S3), but falls short for scenarios S4-S7. GDS-ARM is competitive with GDS(m) for S1-S3, and superior for S4-S7. The same is also true for the middle panel. The right most panel is the most complex situation with smaller mean effect sizes. It is hard for any method to perform well in a complex scenario like this, but GDS-ARM still yields higher power than GDS(m), even though it also has larger error. 

Three additional designs for $n=18$, $m=22$ are considered in the Supplementary Material: two UE($s^2$)-optimal designs \citep{JM14}, and a Bayes D-optimal design \citep[from][]{MW10}. GDS-ARM outperforms GDS(m) and GDS(m+2fi) across scenarios and designs. 
\end{example}

\begin{figure}[hbtp]
	\centering
		\includegraphics[scale=0.6]{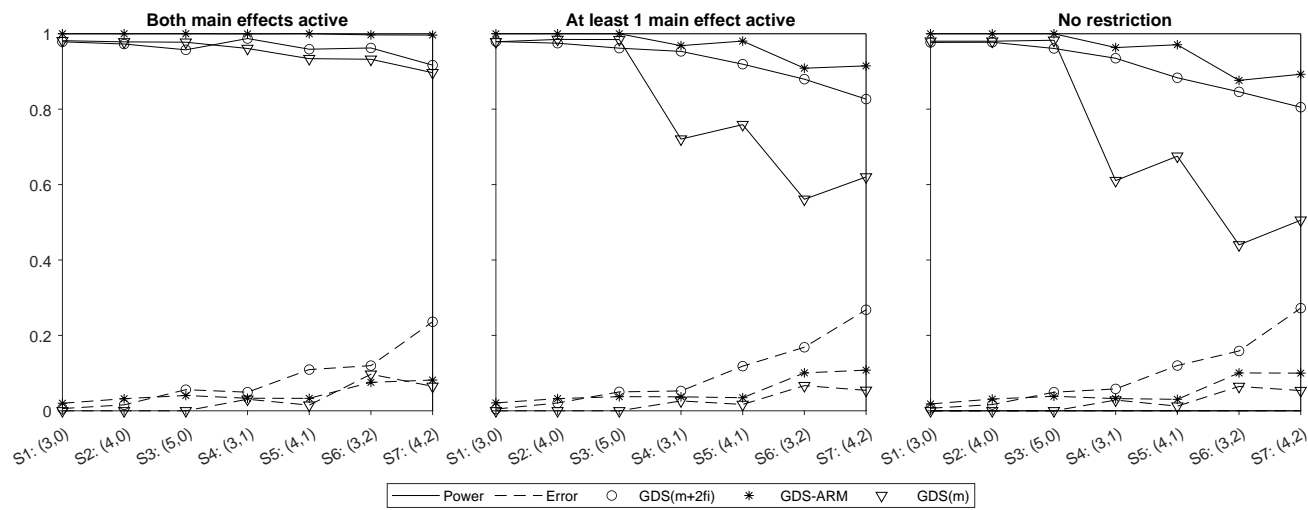}
		\vspace*{-10mm}
	\caption{Average power and error over 1000 iterations for $n=24,m=16$ when coefficients of active effects are generated from a normal distribution with standard deviation 1 and mean effect size of 5. For an active interaction, both main effects are active (left panel), at least one main effect is active (middle panel), or there is no restriction on the main effects (right panel).}
	\label{fig-Main-n24m16}
\end{figure}

\begin{example}\label{exmpln24m16}
For $n=24$ and $m=16$, we use design 20.1 from \cite{schoenetal}. Figure~\ref{fig-Main-n24m16} has coefficients of active effects from $N(5,1)$, however, we consider three different ways for generating models. In the left panel, an interaction between two factors can only be active if both of the corresponding main effects are active (strong heredity). For the middle panel, this restriction is relaxed to at least one of the corresponding main effects is active (weak heredity), while simulation results in the right panel do not impose any such restriction. Designs 20.2a, 20.2b, and 20.3a of \cite{schoenetal} are also studied in the Supplementary Material. The tuning parameters for GDS-ARM are $n{rep}=120$, $nint=24$, $n{top} = 20$, and $pkeep=0.25$. The three panels in Figure~\ref{fig-Main-n24m16} have a very similar performance, especially for GDS-ARM. Figure~\ref{fig-Main-n24m16} shows again that GDS-ARM outperforms GDS(m) and GDS(m+2fi), a conclusion that also holds for the designs studied in the Supplementary Material.

Note that GDS(m+2fi) is not quite as bad for this example as in the previous one, which is related to $n$ being larger and $m$ being smaller in the current example. Examples corresponding to $(n,m)$ = $(20,16)$ and $(24,20)$  in the Supplementary Material also demonstrate this characteristic.
\end{example}

\begin{figure}[hbtp]
	\centering
		\includegraphics[scale=0.8]{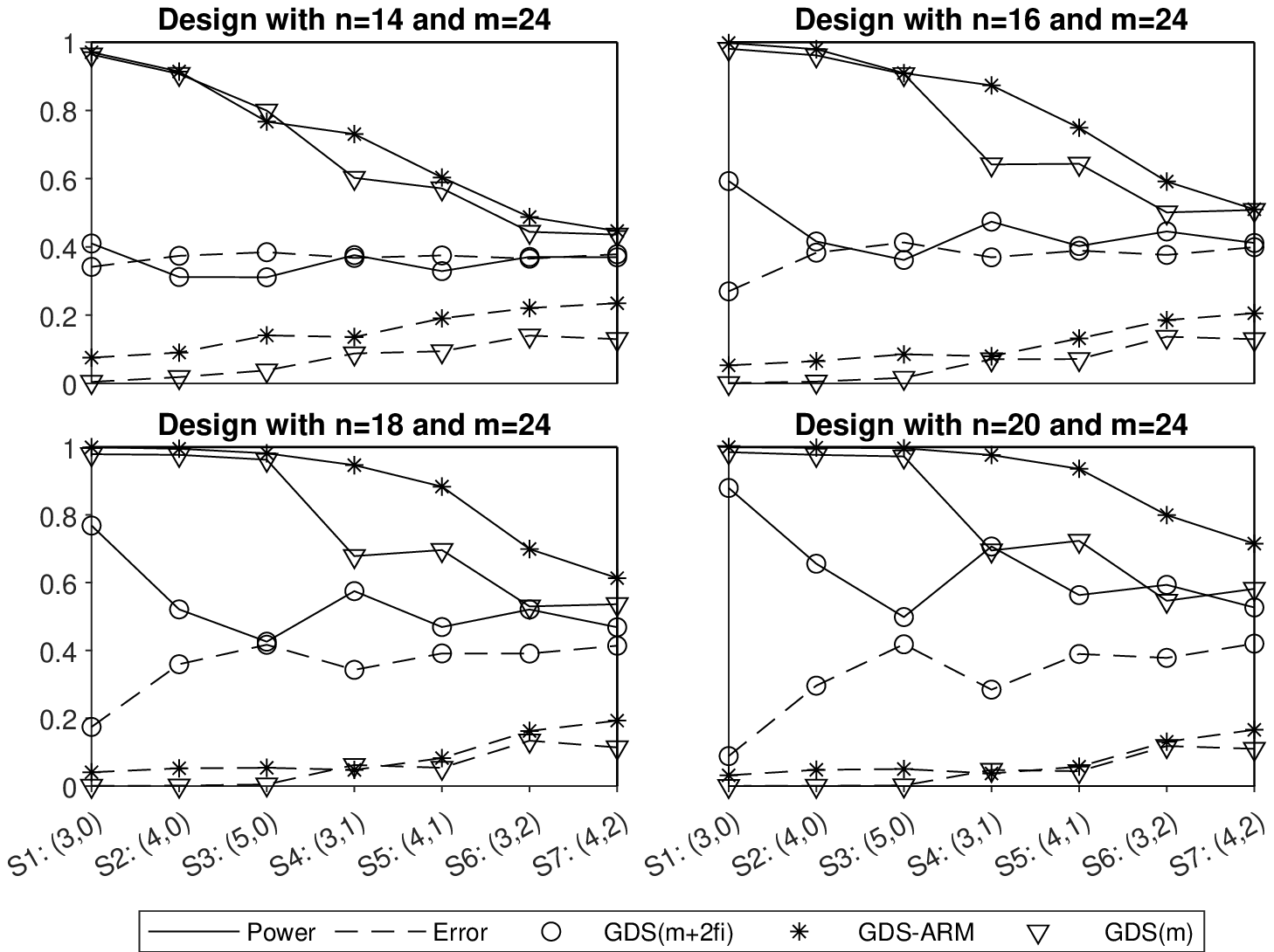}
		\vspace*{-5mm}
	\caption{Average power and error over 1000 iterations for $n=14, 16, 18,$ and $ 20$, $m=24$. The coefficients of active effects are generated from $N(5,1)$ such that at least one of the main-effects corresponding to an active interaction is also active.}\label{fig-Main-n14to20m24}
	\end{figure}

\begin{example}\label{exn14to20m24}
Keeping $m$ fixed at 24, we now consider four $E(s^2)$-optimal designs for $n=14,16,18$, and $20$ respectively. Coefficients of active effects are again drawn from $N(5,1)$ and for any active two-factor interaction at least one of the corresponding main effects must be active. Figure~\ref{fig-Main-n14to20m24} shows that the performance of all methods improves when $n$ increases. While GDS(m) is competitive with GDS-ARM for $n=14,m=24$, GDS-ARM is better in the other three scenarios. We see that for small $n$, none of the methods performs particularly well. But when $n$ comes closer to $m$, GDS-ARM is clearly superior. It is not surprising that, with a very small $n$, it is hard to identify active effects correctly. This latter observation is also visible for the case $(n,m)=(12,16)$ studied in Supplementary Material. 

The tuning parameters for GDS-ARM are according to the recommendations in Table \ref{tab-partuning}. The $E(s^2)$-optimal designs for $n=14, m=24$ is the same as that used in \cite{MW10}, whereas for the other parameters, designs are obtained from \url{https://engineering.purdue.edu/Smartdesigns/twolevel.html.} These designs are also provided in the Supplementary Material.
\end{example}

The Supplementary Material contains many additional examples that lead to observations that are worth mentioning. First we consider different designs for $(n,m)$ = (18,21), (18,22), (12,16), and (24,16). For a given choice of  $n$ and $m$, while most designs perform equivalent, there are times when design choice matters such as for (18,21) and (12,16). For $n=12$ and $m=16$, we studied five designs, one of which (d5) is constructed in \cite{shitang}, with descriptions for constructing the other designs also being provided there. These designs can also be found in the Supplementary Material. Design d5 is meant to allow for the identification of active main-effects without assuming that two-factor interactions are negligible. However, from scenarios S4-S7 in our simulations, it is clear that d5 performs poorly compared to the other designs when there are active two-factor interactions. Results for the design in Example~\ref{exmpln18m22} with $n=18$ and $m=22$ when different models for generating the data are considered are provided in the Supplementary Material. Additionally, results for the design in Example~\ref{exmpln24m16} with $n=24$ and $m=16$ using different mean effect sizes are provided in the Supplementary Material. The conclusions remain the same as described in the Examples~\ref{exmpln24m16} and \ref{exmpln18m22}, respectively. Finally, we consider designs with 24 runs and $m=16, 20, 28,$ and $32$, where consistently GDS-ARM performs better than the other methods.

For almost all the cases we considered, the recommendations for tuning parameters suggested in Table \ref{tab-partuning} work. Performance of the $n=24$ and $m=16$ example for slightly more complicated scenarios are provided in the Supplementary Material. For situations with $n> m$, using a slightly larger value of $nint$ such as $0.4{m\choose 2}$ works a little better than the suggested choice of $0.2{m\choose 2}$, but the choice of $0.2{m \choose 2}$ appears to be quite robust for the values of n and m that we consider. For much larger values of $n$, GDS(m+2fi) is a viable method. For example, for a design with $(n,m)=(94,20)$ \citep{Draguljic2014}, GDS(m+2fi) performs well. Using their scenarios, we studied GDS-ARM. Since the scenarios in \cite{Draguljic2014} tend to have a larger number of active interactions than in our scenarios, we need to select a larger value for $nint$ for GDS-ARM to perform equivalently to GDS(m+2fi). The results in the Supplementary Material for this case use both $nint = 0.2{m\choose 2}$ and $nint = 0.6{m\choose 2}$, and the conclusion is that the latter choice is much better. We reiterate, however, that the focus of the paper is on designs where $n$ is much smaller and where GDS(m+2fi) is a relatively poor screening method. 

The principle of effect heredity was first introduced in \citep{hamadawu92} and later developed by \cite{chipmanetal}. Variable selection methods that incorporates effect heredity include the works by \cite{yuanetal07} and \cite{yuanetal09}. Since the literature suggests that the effect heredity is an important principle to consider, we also considered a modified version of Algorithm \ref{algo1} that embeds (weak) effect heredity in lines 4 and 9 of the algorithm. In line 4, we modify each GDS model by deleting active interaction effects for which neither of the corresponding main effects is active. Similarly, in line 9, after the stepwise model selection, we retain only those interactions that are consistent with the weak heredity principle. This modified version has approximately the same power and a smaller error compared to the original GDS-ARM when the model for generating the data satisfies weak (or strong) heredity. However, for models that do not satisfy this assumption, the modified GDS-ARM can have considerably smaller power and somewhat larger error than the original GDS-ARM. If one firmly believes in weak or strong heredity, one may want to consider the modified version of GDS-ARM. An R package \texttt{GDSARM} which incorporates the GDS-ARM procedure with and without strong and weak heredity is currently submitted to CRAN. Though GDS-ARM is not introduced for computational gains, it is quite fast. For example, For example, for $n=18$, $m=22$, it take about (1/10)th of a second for each GDS run and hence $nrep * (1/10) =  = 24$ secs to run the GDS-ARM on a Desktop with an Intel Xeon CPU @ 3.70GHz processor with 32GB RAM.

\section{Conclusions}
For a complicated process that can be affected by many factors, one should expect that there are some active interactions. If interactions are completely ignored in factor screening, this can lead to erroneous conclusions, both through failing to select some important factors and through incorrectly selecting some factors that are not important. Due to the limited number of runs, a model with only main-effects and all two-factor interactions already becomes extremely complex due to high correlations between model columns. With a nod to the effect hierarchy principle, we must therefore assume that interactions of three or more factors are negligible. Additionally, because of effect sparsity, we focus on situations with a relatively small number of active effects, with more active main-effects than active two-factor interactions.

While GDS is a popular analysis method for screening experiments, neither GDS(m) nor GDS(m+2fi) performs well for a model with main-effects and two-factor interactions, especially when $n$ is relatively small. We proposed a new analysis method, GDS-ARM, which identifies important factors by aggregating results from multiple GDS applications performed on models with different sets of randomly selected interactions. GDS-ARM draws its motivation in part from random forests by using models that contain only some of the available effects, and by identifying important factors after applying GDS on all of these models. Through the simulations and real case studies, we demonstrate that GDS-ARM works well across a range of scenarios, designs, and different values of $n$ and $m$. Our choice of tuning parameters (Table~\ref{tab-partuning}) focuses on the scenarios that have 0 to 2 active interactions. If it is anticipated that more interactions could be active, a different choice of $nint$ could yield better results. Largely, it seems that the choice of design does not matter much. However, as seen in the Supplementary Material for $n=18, m=22$ and $n=12,m=16$, some designs can perform poorly. Further investigations are necessary to identify designs that work best for GDS-ARM. A straightforward generalization of GDS-ARM to factors with three or more levels is possible, however, more work would be needed to tune the parameters for such designs.

\bigskip
\begin{center}
{\large\bf SUPPLEMENTARY MATERIAL}
\end{center}

The Supplementary Material is available and contains information as described in the paper along with the MATLAB codes for GDS-ARM.

\bibliographystyle{Chicago}

\bibliography{literature}

\begin{thebibliography}{}

\bibitem[\protect\citeauthoryear{Bien, Taylor, and Tibshirani}{Bien
  et~al.}{2013}]{bienetal}
Bien, J., J.~Taylor, and R.~Tibshirani (2013).
\newblock A lasso for hierarchical interactions.
\newblock {\em Annals of {S}tatistics\/}~{\em 41\/}(3), 1111--1141.

\bibitem[\protect\citeauthoryear{Booth and Cox}{Booth and Cox}{1962}]{BC62}
Booth, K. H.~V. and D.~R. Cox (1962).
\newblock Some systematic supersaturated designs.
\newblock {\em Technometrics\/}~{\em 4\/}(4), 489--495.

\bibitem[\protect\citeauthoryear{Box and Meyer}{Box and Meyer}{1993}]{boxmeyer}
Box, G. E.~P. and R.~D. Meyer (1993).
\newblock Finding the active factors in fractionated screening experiments.
\newblock {\em Journal of Quality Technology\/}~{\em 25\/}(2), 94--105.

\bibitem[\protect\citeauthoryear{Cand{\`e}s and Tao}{Cand{\`e}s and
  Tao}{2007}]{CandesTao}
Cand{\`e}s, E. and T.~Tao (2007).
\newblock The {D}antzig selector: Statistical estimation when p is much larger
  than n.
\newblock {\em Annals of {S}tatistics\/}~{\em 35\/}(6), 2313--2351.

\bibitem[\protect\citeauthoryear{Chipman, Hamada, and Wu}{Chipman
  et~al.}{1997}]{chipmanetal}
Chipman, H., M.~Hamada, and C.~F.~J. Wu (1997).
\newblock A bayesian variable-selection approach for analyzing designed
  experiments with complex aliasing.
\newblock {\em Technometrics\/}~{\em 39\/}(4), 372--381.

\bibitem[\protect\citeauthoryear{Dopico-Garc{\'\i}a, Valentao, Guerra, Andrade,
  and Seabra}{Dopico-Garc{\'\i}a et~al.}{2007}]{dopico07}
Dopico-Garc{\'\i}a, M., P.~Valentao, L.~Guerra, P.~B. Andrade, and R.~M. Seabra
  (2007).
\newblock Experimental design for extraction and quantification of phenolic
  compounds and organic acids in white “vinho verde” grapes.
\newblock {\em Analytica {C}himica {A}cta\/}~{\em 583\/}(1), 15--22.

\bibitem[\protect\citeauthoryear{Dragulji{\'c}, Woods, Dean, Lewis, and
  Vine}{Dragulji{\'c} et~al.}{2014}]{Draguljic2014}
Dragulji{\'c}, D., D.~C. Woods, A.~M. Dean, S.~M. Lewis, and A.~J.~E. Vine
  (2014).
\newblock Screening strategies in the presence of interactions.
\newblock {\em Technometrics\/}~{\em 56\/}(1), 1--15.

\bibitem[\protect\citeauthoryear{Fan and Li}{Fan and Li}{2001}]{fanli2001}
Fan, J. and R.~Li (2001).
\newblock Variable selection via nonconcave penalized likelihood and its oracle
  properties.
\newblock {\em Journal of the American Statistical Association\/}~{\em
  96\/}(456), 1348--1360.

\bibitem[\protect\citeauthoryear{Georgiou}{Georgiou}{2014}]{georgiou}
Georgiou, S.~D. (2014).
\newblock Supersaturated designs: A review of their construction and analysis.
\newblock {\em Journal of Statistical Planning and Inference\/}~{\em 144},
  92--109.

\bibitem[\protect\citeauthoryear{Hamada and Wu}{Hamada and
  Wu}{1992}]{hamadawu92}
Hamada, M. and C.~F.~J. Wu (1992).
\newblock Analysis of designed experiments with complex aliasing.
\newblock {\em Journal of Quality Technology\/}~{\em 24\/}(3), 130--137.

\bibitem[\protect\citeauthoryear{Hunter, Hodi, and Eagar}{Hunter
  et~al.}{1982}]{Hunteretal}
Hunter, G.~B., F.~S. Hodi, and T.~W. Eagar (1982).
\newblock High cycle fatigue of weld repaired cast {T}i-6{AI}-4{V}.
\newblock {\em Metallurgical Transactions A\/}~{\em 13\/}(9), 1589--1594.

\bibitem[\protect\citeauthoryear{Jones, Lekivetz, Majumdar, Nachtsheim, and
  Stallrich}{Jones et~al.}{2020}]{ryanGOSSD}
Jones, B., R.~Lekivetz, D.~Majumdar, C.~J. Nachtsheim, and J.~W. Stallrich
  (2020).
\newblock Construction, properties, and analysis of group-orthogonal
  supersaturated designs.
\newblock {\em Technometrics\/}~{\em 62\/}(3), 403--414.

\bibitem[\protect\citeauthoryear{Jones, Lin, and Nachtsheim}{Jones
  et~al.}{2008}]{jones2008}
Jones, B.~A., D.~K.~J. Lin, and C.~Nachtsheim (2008).
\newblock Bayesian {D}-optimal supersaturated designs.
\newblock {\em Journal of Statistical Planning and Inference\/}~{\em 138\/}(1),
  86--92.

\bibitem[\protect\citeauthoryear{Jones and Majumdar}{Jones and
  Majumdar}{2014}]{JM14}
Jones, B.~A. and D.~Majumdar (2014).
\newblock Optimal supersaturated designs.
\newblock {\em Journal of the American Statistical Association\/}~{\em
  109\/}(508), 1592--1600.

\bibitem[\protect\citeauthoryear{Li, Sudarsanam, and Frey}{Li
  et~al.}{2006}]{li2006regularities}
Li, X., N.~Sudarsanam, and D.~D. Frey (2006).
\newblock Regularities in data from factorial experiments.
\newblock {\em Complexity\/}~{\em 11\/}(5), 32--45.

\bibitem[\protect\citeauthoryear{Lloyd}{Lloyd}{1982}]{lloyd82}
Lloyd, S.~P. (1982).
\newblock Least squares quantization in {PCM}.
\newblock {\em IEEE transactions on information theory\/}~{\em 28\/}(2),
  129--137.

\bibitem[\protect\citeauthoryear{Marley and Woods}{Marley and
  Woods}{2010}]{MW10}
Marley, C.~J. and D.~C. Woods (2010).
\newblock A comparison of design and model selection methods for supersaturated
  experiments.
\newblock {\em Computational Statistics \& Data Analysis\/}~{\em 54\/}(12),
  3158--3167.

\bibitem[\protect\citeauthoryear{Phoa, Pan, and Xu}{Phoa
  et~al.}{2009}]{phoa2009}
Phoa, F.~K., Y.~H. Pan, and H.~Xu (2009).
\newblock Analysis of supersaturated designs via the {D}antzig selector.
\newblock {\em Journal of Statistical Planning and Inference\/}~{\em 139\/}(7),
  2362--2372.

\bibitem[\protect\citeauthoryear{Phoa, Wong, and Xu}{Phoa
  et~al.}{2009}]{phoachemo}
Phoa, F.~K., W.~K. Wong, and H.~Xu (2009).
\newblock The need of considering the interactions in the analysis of screening
  designs.
\newblock {\em Journal of Chemometrics: A Journal of the Chemometrics
  Society\/}~{\em 23\/}(10), 545--553.

\bibitem[\protect\citeauthoryear{Schoen, Vo-Thanh, and Goos}{Schoen
  et~al.}{2017}]{schoenetal}
Schoen, E., N.~Vo-Thanh, and P.~Goos (2017).
\newblock Two-level orthogonal screening designs with 24, 28, 32, and 36 runs.
\newblock {\em Journal of the American Statistical Association\/}~{\em
  112\/}(519), 1354--1369.

\bibitem[\protect\citeauthoryear{Shi and Tang}{Shi and Tang}{2019}]{shitang}
Shi, C. and B.~Tang (2019).
\newblock Supersaturated designs robust to two-factor interactions.
\newblock {\em Journal of Statistical Planning and Inference\/}~{\em 200},
  119--128.

\bibitem[\protect\citeauthoryear{Tibshirani}{Tibshirani}{1996}]{tibs96}
Tibshirani, R. (1996).
\newblock Regression shrinkage and selection via the lasso.
\newblock {\em Journal of the Royal Statistical Society: Series B
  (Methodological)\/}~{\em 58\/}(1), 267--288.

\bibitem[\protect\citeauthoryear{Vazquez, Schoen, and Goos}{Vazquez
  et~al.}{2020}]{vazquezetal2020}
Vazquez, A.~R., E.~D. Schoen, and P.~Goos (2020).
\newblock A mixed integer optimization approach for model selection in
  screening experiments.
\newblock {\em Journal of Quality Technology\/}, 1--24.

\bibitem[\protect\citeauthoryear{Weese, Edwards, and Smucker}{Weese
  et~al.}{2017}]{mariaweese2017}
Weese, M.~L., D.~J. Edwards, and B.~J. Smucker (2017).
\newblock A criterion for constructing powerful supersaturated designs when
  effect directions are known.
\newblock {\em Journal of Quality Technology\/}~{\em 49\/}(3), 265--277.

\bibitem[\protect\citeauthoryear{Weese, Smucker, and Edwards}{Weese
  et~al.}{2015}]{mariaweese2015}
Weese, M.~L., B.~J. Smucker, and D.~J. Edwards (2015).
\newblock Searching for powerful supersaturated designs.
\newblock {\em Journal of Quality Technology\/}~{\em 47\/}(1), 66--84.

\bibitem[\protect\citeauthoryear{Weese, Stallrich, Smucker, and Edwards}{Weese
  et~al.}{2021}]{mariaweese2020}
Weese, M.~L., J.~W. Stallrich, B.~J. Smucker, and D.~J. Edwards (2021).
\newblock Strategies for supersaturated screening: Group orthogonal and
  constrained var(s) designs.
\newblock {\em Technometrics\/}~{\em 63\/}(4), 443--455.

\bibitem[\protect\citeauthoryear{Westfall, Young, and Lin}{Westfall
  et~al.}{1998}]{westfalletal}
Westfall, P.~H., S.~S. Young, and D.~K.~J. Lin (1998).
\newblock Forward selection error control in the analysis of supersaturated
  designs.
\newblock {\em Statistica Sinica\/}~{\em 8}, 101--117.

\bibitem[\protect\citeauthoryear{Wolters and Bingham}{Wolters and
  Bingham}{2011}]{wolters}
Wolters, M.~A. and D.~Bingham (2011).
\newblock Simulated annealing model search for subset selection in screening
  experiments.
\newblock {\em Technometrics\/}~{\em 53\/}(3), 225--237.

\bibitem[\protect\citeauthoryear{Yuan, Joseph, and Lin}{Yuan
  et~al.}{2007}]{yuanetal07}
Yuan, M., V.~R. Joseph, and Y.~Lin (2007).
\newblock An efficient variable selection approach for analyzing designed
  experiments.
\newblock {\em Technometrics\/}~{\em 49\/}(4), 430--439.

\bibitem[\protect\citeauthoryear{Yuan, Joseph, and Zou}{Yuan
  et~al.}{2009}]{yuanetal09}
Yuan, M., V.~R. Joseph, and H.~Zou (2009).
\newblock Structured variable selection and estimation.
\newblock {\em The Annals of Applied Statistics\/}~{\em 3\/}(4), 1738--1757.

\end{thebibliography}
\end{document}